\documentclass{article}

\usepackage[margin=1in]{geometry}

\usepackage{natbib}
\usepackage{algorithm}
\usepackage{algpseudocode}
\usepackage{hyperref}
\usepackage{graphicx}
\usepackage{amsmath}
\usepackage[toc,page]{appendix}

\author{Dr.~Devan Becker, Wilfrid Laurier University}
\date{}

\title{UnMuted: Defining SARS-CoV-2 Lineages According to Temporally Consistent Mutation Clusters in Wastewater Samples}

\begin{document}
\maketitle

\begin{abstract}
    SARS-CoV-2 lineages are defined according to placement in a phylogenetic tree, but approximated by a list of mutations based on sequences collected from clinical sampling. Wastewater lineage abundance is generally found under the assumption that the mutation frequency is approximately equal to the sum of the abundances of the lineages to which it belongs. By leveraging numerous samples collected over time, I am able to estimate the temporal trends of the abundance of lineages as well as the definitions of those lineages.

    This is accomplished by assuming that collections of mutations that appear together over time constitute lineages, then estimating the proportions as before. Three main models are considered: Two that incorporate an explicit temporal trend with different constraints on the abundances, and one that does not estimate a temporal component. It is found that estimated lineage definitions correspond to known lineage definitions with matching temporal trends for the lineage abundances, despite having no information from clinical samples. I refer to this set of methods as ``UnMuted'' since the mutations are allowed to speak for themselves.
\end{abstract}

\section{Introduction}

The SARS-CoV-2 pandemic has clearly demonstrated the efficacy of wastewater-based epidemiology.
Many recent studies have used wastewater to predict case counts based on the concentration of SARS-CoV-2 in wastewater, detect current lineages of concern, and measure their relative abundance.
Wastewater has also proven useful in detection of cryptic lineages and cryptic mutations.

SARS-CoV-2 RNA molecules will break down in wastewater, making it impossible to reliably sequence entire genomes \citep{linAssessingMultiplexTiling2021}.
Instead of whole genomes, \citet{janssenLessonsLearned2Years2023} have found that approximately 400 base pairs (as implemented in the ARTIC protocol \citep{quickMultiplexPCRMethod2017,quickNCoV2019SequencingProtocol2020}) is an appropriate balance between recoverable genomes and efficiency of information gain.
This means that we do not obtain sequences long enough to determine the SARS-CoV-2 lineage to which they belong (\emph{e.g.}, using NextClade \citet{hadfieldNextstrainRealtimeTracking2018} or PANGO \citet{otoolePangoLineageDesignation2022} definitions).
We instead obtain a collection of mutations, how often they were observed (count), and how often their position on the genome was observed (coverage/depth; \emph{i.e.}, the number of times the mutation could have been observed).

To determine lineages, we first must define what a lineage is.
NextClade \citep{aksamentovNextcladeCladeAssignment2021} and PANGO \citep{rambautDynamicNomenclatureProposal2020} both assign lineages according to placement on a phylogenetic tree built from clinical sequences; lineages are \textit{not} defined according to a set of mutations.
Wastewater genomic sampling results in a set of mutations and thus requires lineage definitions as a list of mutations.
These definitions are often found by querying clinical sequences to see how often a mutation was observed within a given lineage compared to how often it was observed in other lineages, thus creating an alternate definition of a lineage based on it's ``representative'' mutations \citep[see, \emph{e.g.},][]{ellmenLearningNovelSARSCoV22024,swiftCommunitylevelSARSCoV2Sequence2021,karthikeyanWastewaterSequencingReveals2022,valierisMixtureModelDetermining2022}.

Given lists of mutations which define a lineage, estimations of relative abundance of the lineages proceeds according to the following principle: The frequency of a mutation should be equal to the sum of the proportions of each lineage for which it is present.
For example, let $f_{m_1}$ be the frequency (count divided by coverage) of mutation ${m_1}$.
Let $\rho_X$, $\rho_Y$, and $\rho_Z$ be the relative abundance (proportion) of lineages $X$, $Y$, and $Z$ in the population, respectively (I will always use ``frequency'' for mutations and ``proportion'' or ``abundance'' for lineages).
If mutation ${m_1}$ is present in lineages $X$ and $Z$ but not $Y$, then we posit that $f_{m_1} \approx 1\rho_X + 0\rho_Y + 1\rho_Z$, that is, the frequency of mutation $1$ is approximately equal to the sum of proportions of lineages to which it belongs.
This assumption has been imposed by various authors, including: \citet{ellmenAlcovEstimatingVariant2021,karthikeyanWastewaterSequencingReveals2022,valierisMixtureModelDetermining2022,schumannCOVID19InfectionDynamics2021}.

As is implied by the equation, almost all estimators for the relative abundance are based on a linear modelling or generalized linear modelling framework.
The dependent variable is $f_i$, the independent variables represented as 0s or 1s depending on whether mutation $i$ is present in the lineage $j$, and $\rho_j$ taking the place of the coefficients.
Estimation procedures attempt to impose constraints to ensure that the the sum of the abundances is less than or equal to 1 and all of the abundances are between 0 and 1, with some of the estimators incorporating the coverage to ensure better estimates of the standard errors.
These estimators assume a known definition of lineages as a list of mutations, although we know such definitions to be imperfect.

In this work, I leverage repeated wastewater samples to get a temporally varying abundance estimate of each lineage while simultaneously estimating the definitions of those lineages.
Wastewater mutation data are very noisy, so the temporal model smooths out some of the noise, with clusters defined accooding to the temporal pattern that best suits a mutation.
I accomplish this with the following assumption (which replaces the assumption of perfectly known lineage definitions): lineages are defined according to clusters of mutations that persist over time.
My definitions of lineages, then, are lists of mutations that tend to appear together consistently in wastewater (rather than in clinical sequences).
This removes the dependence on clinical sampling, which is costly, subject to bias in the choice of whose infection to sequence and requires many more sequences than we currently obtain in order to get a complete estimate of the genetic diversity of SARS-CoV2.

Other authors, including \citet{ellmenLearningNovelSARSCoV22024} and \citet{zhuangEarlyDetectionNovel2024}, also leverage multiple samples to determine lineage definitions from wastewater alone.
\citet{ellmenLearningNovelSARSCoV22024} use non-negative matrix factorization to find patterns that exist in the observed frequencies of mutations.
\citet{zhuangEarlyDetectionNovel2024} use independent component analysis (a form of blind source separation) to separate mutation frequencies into a collection of signals.
They then take the largest signals from each source as evidence of presence of a mutation.
Further post-processing is performed to determine the relative abundance of each estimated signal.
Both of these approaches require an \textit{a priori} choice of the number of lineages (clusters), but there are well-known diagnostics to help with this choice.

In this work, I leverage the explicit temporal structuring of the samples to develop estimates of the proportions of lineages, which are estimated as clusters that are stable over time.
This estimates a smooth temporal component, which differentiates the method from non-negative matrix factorization.
The estimates of the lineages are part of the estimation procedure, rather than based on post-processing, and are not assumed to be independent, which differentiates this method from independent component analysis.

\section{Methods}

The following model for the relative abundance within a wastewater sample was independently developed by myself and \citet{valierisMixtureModelDetermining2022}.
My version is available as an R package called \texttt{provoc}, which is short for \textbf{Pro}portions of \textbf{V}ariants \textbf{o}f \textbf{C}oncern.
For the purposes of this work, I include the subscript $t$ to denote the discrete time point that the count of mutation $i$ was observed, labelled $c_{i,t}$, the proportion of lineage $j$ at that time point, $\rho_{j,t}$, and the coverage for that mutation at that time, $D_{i,t}$ ($D$ for ``Depth'').
In matrix notation with $N$ mutations and $T$ time points, $c$ is an $N\times T$ matrix (note that this implies that we know the count of each mutation at every time point; there is no missing data, but time points may be irregular).
I use $\underline\rho_t$ to represent the vector of lineage proportions at time $t$ and $\underline\rho_j$ to be the vector of proportions of lineage $j$ over time, which are, respectively, the rows and columns of the matrix $\rho_{N\times J}$ ($J$ being the total number of lineages).
Denote the matrix of lineage definitions $Z$, with binary entries (or possibly entries in the unit interval), columns $Z_{\cdot, j}$ defining each lineage and rows $Z_{i,\cdot}$ defining which lineages mutation $i$ belongs to.
I make the simplifying assumption that coverage $D$ is fixed and that the count $c$ follows a binomial GLM (Generalized Linear Model) as follows:
$$
c_{i,t} \sim \textrm{Binom}(\texttt{prob} = Z_{i,\cdot}\underline\rho_t, \texttt{size} = D_{i,t})
$$
This is a multivariate binomial GLM with the identity link function where all observations have the same covariates but different coefficients.
This is estimated via constrained optimization to ensure that the constraints are satisfied.

I seek to incorporate the dependence structure of $f_{i,t} = Z_{i,\cdot}\underline\rho_t$, i.e. the temporal dynamics of the frequency of a mutation.
Since $Z$ is not subscripted by time, this necessitates a temporal model for $\underline\rho_j$.
Special considerations should be made to ensure that $f_{i,t}\in[0,1]$, \textit{i.e.} $\rho_{j,t}\in[0,1]$, and $\sum_{j=1}^J\underline\rho_j\le 1$.
Note that the constraint on the sum of the parameters is an inequality.
A sum-to-one constraint would assume that the lineages in the model are complete, and no other lineages might exist in the data.
A sum-to-one-or-less constraint allows for the case where there are lineages missing from the provided definitions.

Estimation of $Z$ requires careful consideration.
In a basic clustering situation, one might require that each observation belongs to exactly one cluster, i.e. each mutation belongs to exactly one lineage.
This is clearly not appropriate for SARS-CoV-2 lineages.
Instead, this is a clustering problem where observations can be assigned to multiple clusters based on minimizing a distance based on the sum (not average) of the members of the clusters.
The estimation procedure for assigning mutations into (possibly multiple) lineages is defined below.

\subsection{Non-negative Matrix Factorization (NMF)}

The problem of estimating the lineage definitions at the same time as estimating the proportions of those lineages, especially when the mutation frequencies are a sum of the lineage proportions, can be formulated as a matrix factorization problem.
We can define $Y$ as a matrix with columns representing each sample and rows representing mutations, so that $Y_{m, t}$ is the frequency of mutation $m$ in sample $t$ (note that this does not incorporate the read depth; a frequency of 20/40 is given equal weight to a frequency of 2000/4000).
After choosing $r$, the number of lineages (clusters) to search for, the lineage definition $Z$ is an $N\times R$ matrix of 0s and 1s, while the $R \times T$ coefficients matrix $G$ represents the proportion of each lineage at each time.
The problem can be written as decomposing $Y$ into the matrix product $ZG$, $Y \approx ZG$, where $Y$, $Z$, and $G$ are all non-negative valued matrices, and fit using standard Non-negative Matrix Factorization (NMF) algorithms.

This approach has been previously used by \citet{ellmenLearningNovelSARSCoV22024} for similar data, and I provide another application of this approach here.
Note that this does not apply any restrictions on the $Z$ or $G$ matrices, meaning that the entries in $Z$ are not restricted to be binary (or even in the unit interval), and the estimated prototypes $G$ are not restricted to be similar for observations that are close in time.
However, this does result in an additive relationship with the estimated lineages, thus allowing lineage definitions to overlap.

\subsection{Binomial NMF}

In this section I introduce a version of NMF that uses a Binomial error structure and is fit via Bayesian methods.
The goal is the same as before: we seek to factorize an $n\times t$ matrix $Y$ into $Z_{n\times r}$ and $G_{r\times t}$, but now the matrix $Y$ is the unobserved probability of a success in a binomial distribution with observed count $C$ and observed read depth $D$.
Again, $Z$ represents the estimated lineage definitions (with entries 1 if a lineage contains a particular mutation and 0 otherwise) and $G$ represents the proportion of each lineage at each time point, with $R$ being the number of bases (non-exclusive clusters) to estimate.
It is important to note that the estimates of the lineage definitions, as well as the coefficients, are invariant to permutations of the rows and/or columns of $Y$; that is, this method does not impose any structure on the temporal component.

This is fit under a Bayesian paradigm with the following likelihood and priors:
\begin{align*}
C_{i,t} &\sim \text{Binom}(\texttt{prob} = (ZG)_{i,t}, \texttt{size} = D_{i,t})\\
Z_{i,r} &\sim \text{Bernoulli}(0.5)\\
G^*_{\cdot, t} &\sim \text{Dirichlet}(\underline 1)\\
G^I_{j} &\sim \text{Bernoulli}(0.5)\\
G_{j, t} &= G^*_{j, t} * G^I_{j}
\end{align*}
The Bayesian models in this paper are fit using the NIMBLE library in the R programming language \citep{rcoreteamLanguageEnvironmentStatistical2024,devalpineNIMBLEMCMCParticle2022}.

For a particular time point $t$, $G^*_{\cdot, t}$ follows a Dirichlet distribution (with a prior parameter that is a length $J$ vector of ones) in order to ensure that the coefficients are non-negative and sum to one.

Note that this method does not allow for missing lineages; the lineages estimated are required to account for all lineages in the data.
The setup also includes a switch to ``turn off'' specific lineages, essentially allowing the method to choose $R$, the number of lineages in the data (if $G^I_r = 0$, then the $r$th lineage is not present in the data).
Also note that, since the Dirichlet distribution ensures that $\sum_j G_{j,t} = 1$ and all entries of $Z$ are either 0 or 1, this ensures that $0 \le (ZG)_{j,t} \le 1$ for all entries and $\sum_{j} (ZG)_{j, t}\le 1$ for all $t$, thus $ZG$ is always a valid proportion and there is no need for a transformation such as logit or probit.

\subsection{Temporal Binomial NMF with Sum-to-One-or-Less (TBNMF$\le$1)}

The NMF model described above can be improved upon by incorporating an explicit temporal structure.
Rather than $G$ following a Dirichlet distribution for a particular day, a B-spline basis expansion can be used to estimate a smooth temporal trend for a particular lineage.
With this setup there is no simple mechanism by which the estimates can be guaranteed to be valid proportions, so transformations to the unit interval are required.

The likelihood is the same as above, but the matrix $G$ is estimated as a basis spline estimate for the time series of each lineage, transformed to be valid proportions (with a sum of one \emph{or less}).
A set of B-spline basis functions $B = [b_1(t), b_2(t), ..., b_M(t)]$ is defined, where the abundance of each lineage will be estimated as $G_{j, t} = \sum_{m = 1}^M\phi^*_(j, m)b_m(t)$.
The basis coefficients are assumed to have a uniform prior distribution on the unit interval, as there is no reason for an individual basis coefficient to be negative nor greater than one.
Rather than ``turning off'' a lineage, individual coefficients for basis splines for a particular lineages can be set to zero, using an indicator function as before.

Assuming that $b$ is a matrix with columns $b_m(t)$,
\begin{align*}
\phi^*_{j, m} &\sim \text{Uniform}(0, 1)\\
\phi^I_{j} &\sim \text{Bernoulli}(0.5)\\
\phi_{j, m} &= \phi^*_{j, m} * \phi^I_{j}\\
G^* &= B\phi\\
G_{\cdot, t} &= f(G^*_{cdot, t})
\end{align*}
where $f(\cdot)$ is a vector-valued function that I refer to as ``fluffmaxing'':
$$
f(\underline a) = \dfrac{\underline a}{\sum_i a_i) * I(\sum_ia_i > 1) + I(\sum_ia_i) \le 1)}
$$
This function will divide by the sum of the input if the sum is larger than 1 (thus forcing the output to have a sum of 1), or return the original value if the sum is less than one.
This is appropriate for the present situation since we want to allow for the case where some lineages are not searched for in the wastewater data.

The Uniform(0,1) distribution was used for $\phi^*$ to keep the spline estimates low.
The spline estimates are still able to grow larger than one, but the MCMC algorithm will not be able to explore too far outside of the unit interval.
Note the inclusion of $\phi^I_{j}$, which is an indicator variable.
When this variable is 0, then lineage $j$ is set to 0 at all time points.
This allows us to over-specify the number of lineages, then allow the estimation routine to reduce the number to an appropriate amount.

The prior for $Z$ and the likelihood are the same as before.
Note that this formulation of the model appears more complicated, but converges faster than the binomial NMF due to the restrictions on the parameters which results in a smaller number of effective parameters.

\subsection{Temporal Binomial NMF with Sum-to-One-or-Less (TBNMF=1)}

The model formulation above goes to great lengths to ensure that the sum of the coefficients at time $t$ is 1 \emph{or less}.
This restriction is important for the model in which lineages must be pre-specified since the pre-specified lineages may be incomplete, and thus we are missing a summand in the sum of proportions.
However, for the present model, we can return to the sum-to-exactly-one constraint by re-introducing the Dirichlet distribution, but with the spline-based priors.

The estimation of $G$ is achieved with the following priors:
\begin{align*}
\phi^*_{j, m} &\sim \text{Exponential}(0, 1)\\
\phi_{j, m} &= \phi^*_{j, m} * \phi^I_{j, m}\\
G^* &= b\phi\\
G_{\cdot, t} &\sim \text{Dirichlet}(G^*_{\cdot, t})
\end{align*}

This has the added consequence of allowing extra variation in the mutation frequencies, as the Dirichlet distribution has its own variance.
Note that the prior on $\phi^*$ is now exponential, since values larger than 1 are allowed in the Dirichlet distribution (the sum will still be one, but the variance will be lower with larger values).

\section{Applications}

\subsection{Data Preparation}

For demonstration, I use publicly available data deposited in NCBI BioProject {PRJNA1088471}, which includes samples from several wastewater treatment plants (WWTPs) in Toronto, Ontario, Canada as well as sites at Pearson Internation Airport.
Since we are focused on the temporal aspect of the data, we focus on the data from the Highland Creek WWTP.
This plant was chosen because it is a WWTP (airport wastewater will be inherently more variable) and mostly has equally spaced sampling dates (82 unique sampling dates total, almost always separated by 7 days).
See \cite{overtonGenomicSurveillanceCanadian2024a} for a detailed description of the genetic sequencing procedure.
These data were processed using the GromStole pipeline (\url{https://github.com/PoonLab/GromStole}) to produce a list of mutations relative to the Wuhan-1 reference genome, the number of times the mutations were observed (``count''), and the number of reads that aligned to any given position on the genome (``coverage'' or ``depth'').

\begin{figure}[ht!]
\includegraphics[width=0.98\textwidth]{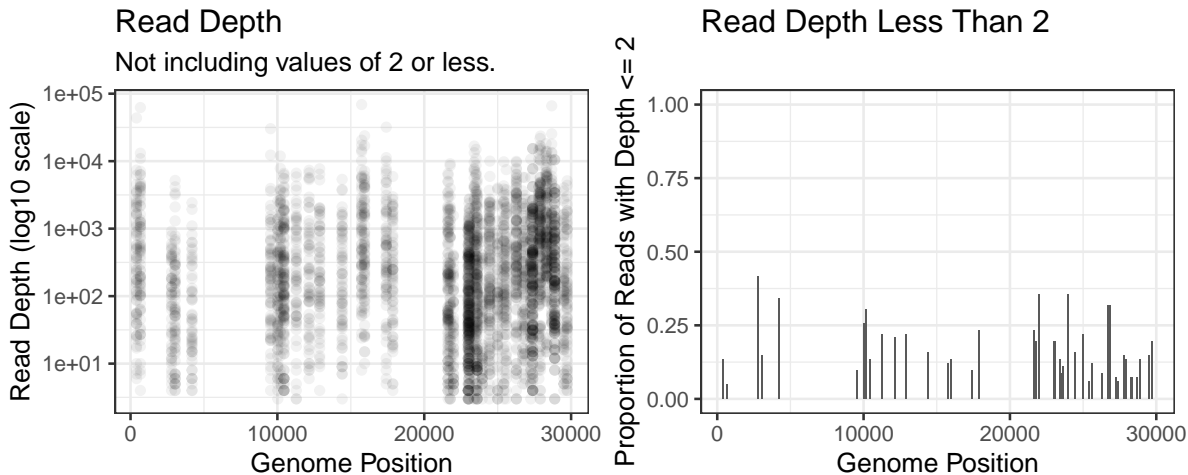}
\caption{\label{fig:coverage}\textbf{Left}: The read depth (on the log scale) of each mutation at each position on the genome, filtered to only include values of 3 and aboved. Points are partially transparent to highlight areas of higher density. Since we require observations of each mutation at each time point, many mutations have a coverage of 0 (which were replaced with 1 in the data, resulting in a frequency of 0/1). \textbf{Right}: The proportion of observations at each position that had a read depth of 2 or fewer. These observations would normally be too low to include, but are necessary to have observations of each mutation on each sampling day.}
\end{figure}

The mutation data were further processed as follows.
Initially, only mutations with a read depth of at least 40 were considered in any given sample.
Among these mutations, I set a threshold $d$ for each mutation over time: a mutation had to have a frequency of at least 10\% for $d$ time points in the study \textbf{and} a frequency less than 90\% for $d$ time points in the study, with $d$ investigated at values of 10, 15, or 20.
This ensures that I include mutations with non-trivial temporal dynamics.
Once the set of suitable mutations was determined, I required observations of these mutations at all time points in the study, regardless of their read depth (to avoid dividing by zero, read depths of 0 were replaced with a 1).
This resulted in 88 unique mutations for $d=10$, 72 for $d = 15$, and 57 for $d = 20$.

In the event that several samples were collected on the same day, I combined the samples by adding the counts together and adding the coverages together, then re-calculating the frequency.
This is equivalent to simply concatenating the short read files, and assumes that each sample was performed the same (in all data analysed, repeat samples all used the same instruments with the same settings).

\subsection{Independent Binomial GLMs (``ProVoC'')}

The model detailed at the start of the Methods section can be fit independently for each sample.
This will be instructive for interpreting the results of the clustering algorithm (noting that the estimates are not a ``ground truth'' since the lineage definitions are imperfect).
The ``ProVoC'' model is a binomial GLM with the identity link function, a linear predictor based on binary covariates defined according to the lineage definitions in \url{https://github.com/cov-lineages/constellations}, and restrictions on the coefficients to ensure that estimated proportions are positive and sum-to-less-than-one.
This model is similar to the one posited by \citet{valierisMixtureModelDetermining2022}; this application was developed independently by myself and implemented in \url{https://github/DASL-Lab/provoc}.

\begin{figure}
    \centering
    \includegraphics[width=0.99\textwidth]{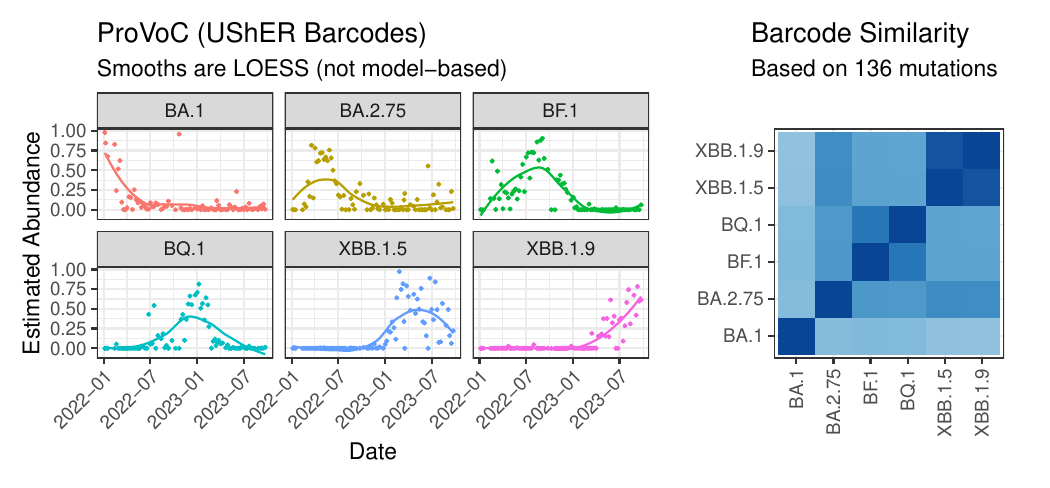}
    \caption{\label{fig:provoc}\textbf{Left}: Proportions of each lineage as estimated by ProVoC, a binomial GLM with constraints fit to each sample individually. \textbf{Right}: Jaccard Similarity of the UShER Barcode definitions from the Freyja variant caller.}
\end{figure}

To define the lineages, I used the ``barcodes'' file supplied by the Freyja variant caller \citep{karthikeyanWastewaterSequencingReveals2022}.
The plot on the right of Figure \ref{fig:provoc} demonstrates the Jaccard similarity of these lineages.
These definitions consist of a list of mutations, with many mutations being part of more than one lineage (in fact, there is no pair of lineages which do \emph{not} share at least some mutations).
From the plot on the right in Figure \ref{fig:provoc}, we can see that BF.1 and BQ.1 have quite similar definitions, and BA.2.75, XBB.1.5, and XBB.1.9 have similar definitions.

Results are shown in Figure \ref{fig:provoc}.
The smooth lines are \emph{not} based on any model-based temporal smoothing; they are merely LOESS smooths of the data without any incorporation of the variance of the results and are included purely for visual aid.
Note that these results are not a ``gold standard'' or a ``ground truth'', but rather are shown to demonstrate that the given location saw a transition between numerous lineages throughout the study.
To fit this model I had to make decisions about which lineages might be present, and different decisions would lead to different results.
Also note that the definitions of the XBB lineages differ by only a handful of mutations (see the Jaccard similary of the UShER barcodes in Figure \ref{fig:all-bases}).

The results show a clear set of ``waves'' of each of the six lineages that I chose to include.
BA.1 had high abundance at the start, then decreased to approximately 10\% by July 2022 and stayed at this abundance until January 2023.
BA.2.75 took over after BA.1, but never dropped to 0 according to the LOESS smooth of the results.
The next waves were BF.1, then BQ.1, then XBB.1.5, with the XBB.1.9 wave still ongoing at the end of the data being analyzed.
These observations are concordant with the known lineages from clinical sampling in Toronto during this time period.

\subsection{Non-negative Matrix Factorization (NMF)}

There is no universally applicable method for choosing the rank (\textit{i.e.}, number of estimated lineages) of the factorization, $R$.
Here, values of $R$ between 4 and 12 were investigated.

\begin{figure}[ht!]
    \centering
    \includegraphics[width=0.99\textwidth]{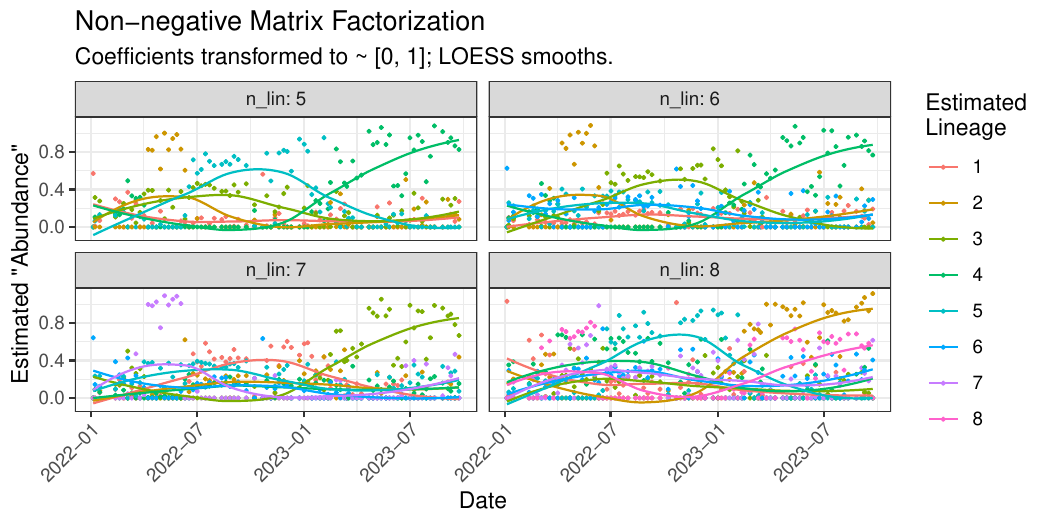}
    \caption{\label{fig:nmf}Coefficients representing the prevalence of each estimated lineage from NMF, fit separately to each location. The lineage names are arbitrary; they are not based on placement in a phylogenetic tree as defined by clinical cases. In all cases, the results approximately match the results from when the lineages were known from clinical cases.}
\end{figure}

The results were post-processed so that the estimated lineage definitions were (almost) valid proportions by dividing each entry in $Z$ by the 99th percentile of all values (the 99th percentile was chosen because it produced reasonable estimates for the coefficients as proportions).
The resultant coefficients are shown in Figure \ref{fig:nmf}.

Note that this transformation requires carefull consideration if it is used to determine the coefficients and definitions in practice.
Ideally, a transformation of $Z\times c$ would be used to ensure that the sum-to-one constraint holds for each time point.
This transformation would need a corresponding quantity $c'$ to satisfy $(Z\times c)\times (c'\times G) = Z\times G$, and such a value $c'$ is not guaranteed to exist.
See \citet{ellmenLearningNovelSARSCoV22024} for the post-processing methods they used in order to ensure valid proportions and the corresponding definitions.
In the present work, we are using NMF as a partial step towards more approporiate methods, and thus do not consider further post-processing.

The summary statistics for the NMF runs with ranks  of 2 to 20 are shown in Figure \ref{fig:nmf_rank} in the Appendix.
These plots do not suggest a clear ``best'' rank.
The cophenetic score should be chosen as the highest value before a significant drop, and ranks of 3, 5, 7 to 10, and possibly 14 appear to be suitable ranks.
The dispersion suggests that ranks of 5 to 9 might be appropriate.
There is no clear ``elbow'' in the Explained Variance (\texttt{evar}), Residuals, or RSS.
A higher silhoutte score before a decrease is another way to choose the rank; the scores for ``Basis'' and ``Coefficients'' appear to have higher values (relative to the surrounding ranks) at ranks 5, 6, and 7 before they decrease, although it is not a steep decrease.
From this ad-hoc investigation, I have run the other models with ranks of 4 to 11.

The results for 6 lineages is mostly concordant with the results from ProVoC (where 6 known lineages were found to adequately explain the results).
The plots become difficult to read with too many lineages, so the remaining plots are shown in the appendix (Figure \ref{fig:nmf_all}.

Note that this algorithm does not have information about the clinical sequences; no information about the PANGO nomenclature is included.
The mutations were identified independent of the definitions of lineages in PANGO.
As such, the lineage names are given as numbers.

The lineage definitions were chosen according to a clustering of mutations with similar frequencies across samples.
There is nothing in this algorithm that constrains the coefficients over time, nor does it constrain the lineage definitions to be 0s or 1s, nor does it constrain the coefficients to be between 0 and 1 (as is clearly seen from the y-axis in Figure \ref{fig:nmf}).
In fact, the algorithm doesn't even incorporate temporal ordering when identifying clusters.
Despite this, the algorithm still identifies a temporal trend of the lineages that were found with ProVoC.

\subsection{Binomial NMF}

\begin{figure}[ht!]
    \centering
    \includegraphics[width=0.99\textwidth]{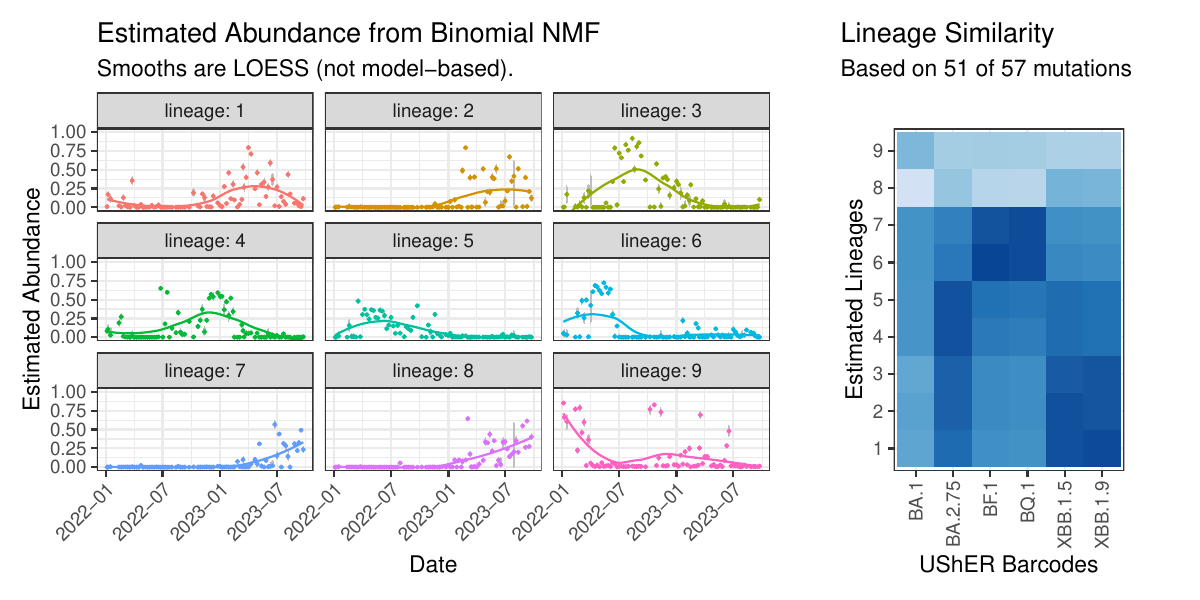}
    \caption{\label{fig:bbnmf}\textbf{Left}: Results from the Binomial NMF algorithm. Again, the lineage names are arbitrary in each run; the names cannot even be compared across plots (no post-processing was done to align the labels between locations). \textbf{Right}: Jaccard similirity of the estimated lineages with the definitions given in UShER Barcodes, with darker colours indicating higher similirity. Jaccard similarity uses the intersection of the lineage definitions, thus there are mutations in the estimated definitions that were not used in the similarity calculation and mutations in the known definitions that were not used.}
\end{figure}

The model was fit as described in the methods, using the priors described there.
Unlike with NMF, all proportions estimated from a given sample will sum to 1 and the basis functions consist solely of 0s and 1s.
The model includes a method for removing lineages, but in my testing it does not appear to have reliably set many lineages to 0.
To decide the number of lineages, I fit the model with 4 to 13 lineages and chose the model that minimized the WAIC \citep{watanabeAsymptoticEquivalenceBayes2010} before an increase.
The results of this are shown in Figure \ref{fig:waic}.
This resulted in an optimal choice of 10 lineages, however, one of those lineages was set to 0 (see Figure \ref{fig:bbnmf_rank}) and thus I chose 9 lineages.
The number of basis functions was set as 10 \textit{a priori} according to visual inspection of the results for 8, 10, 12, and 14 basis functions.

Results are shown in \ref{fig:bbnmf}, with lineages indicated as numbers since they do not correspond to placement in a phylogenetic tree.
The numbers were chosen according to the Jaccard similarity of each estimated lineage to the UShER barcode definition of XBB.1.9, with lineage ``1'' being the most similar.

From Figure \ref{fig:bbnmf}, we can see that this algorithm has split some of the known lineages into multiple lineages.
Estimated lineages labelled 1, 2, and 3 were all similar to XBB.1.9 (as expected by my choice of ordering), but were also very similar to XBB.1.5 and BA.2.75.
Lineage 9 was ``most'' related to BA.1 according to the Jaccard similarities, and the plot of abundance over time supports this.
However, the similarity is quite low relative to other similarities found in the estimated lineages.
As noted in Figure \ref{fig:provoc}, BF.1 and BQ.1 are similar to each other, so it is not surprising that the most similar estimated lineages (6 and 7) are similar to each other as well as both BF.1 and BQ.1.
BA.2.75 is the ancestral lineage (or source of recombination) for every lineage considered except for BA.1, so it is not surprising that the mutations in this lineage were spread across many of the estimated lineages.
For a plot of the Jaccard similarities amongst the estimated lineages, see Figure \ref{fig:all-bases}.

Estimated lineage 8 was not particularly similar to any of the lineages that I used from the barcodes file.
This may indicate a superfluous lineage, or it might be an indication that I did not specify enough lineages in my initial ProVoC analysis.
This estimated lineage only gained abundance toward the end of the sampling period, so this might indicate the presence of a cryptic lineage not yet captured by known clinical definitions.

\subsection{Temporal Binomial NMF with Sum-to-One-or-Less (TBNMF$\le$1)}

Unique to this method is the fact that the lines shown on the plot are the estimates themselves, rather than a LOESS smooth of the estimated coefficients.
The estimates are inherently temporal, which imposes further structure on the estimated lineages: the mutations that are grouped together must all have similar frequencies \emph{on neighbouring dates}.

\begin{figure}[ht!]
    \centering
\includegraphics[width=0.99\textwidth]{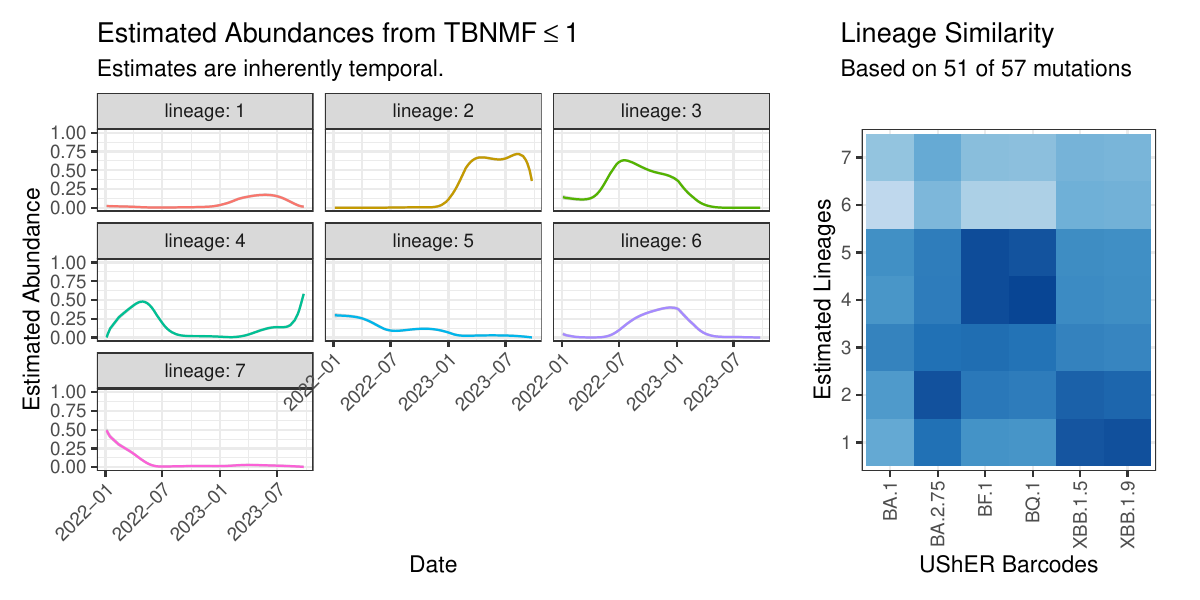}
\caption{\label{fig:unmuted}\textbf{Left}: Estimated temporal smoothers for abundances of lineages shown on the right. \textbf{Right}: Jaccard similarity of each estimated lineage with each lineage defined from UShER barcodes (darker colours indicate higher similarity). Lineage numbers are chosen in order of similarity to XBB.1.9.}
\end{figure}

This method imposes the assumption that the coefficients must sum to one \emph{or less}.
This allows for the possibility of more lineages than specified in the data, or the possibility that the data processing step missed important mutations that might otherwise provide information on potential lineages.

According to the WAIC (Figure \ref{fig:waic}), either 6, 7, or 11 lineages were optimal.
With $R=11$,there was a lineage whose coefficient was always 0.
Looking at $R=10$, there were two other lineages that had a coefficient that was set to 0. 
For this reason, I demonstrate the estimates with 7 estimated lineages ($R = 7$), which was the other optimum number.
Note that the choice of rank in any NMF-like algorithm is going to be subject to these sorts of subjective decisions.
Again, the number of basis functions was set as 10.

Results for the sum-to-one-or-less constraint are shown in Figure \ref{fig:unmuted}.
As in BNMF, lineages estimated here are a combination of the lineages defined by UShER barcodes.
No lineage is particularly similar to BA.1, but lineages 5 and 7 match its temporal trends despite being comprised of different mutations.
Most estimated lineages contain mutations from BA.2.75, which is the ancestral lineage or base for recombinant to all but BA.1.
Again we see several estimated lineages (6, 7, and 8) that are very similar to both BF.1 and BQ.1.
Finally, lineages 6 and 7 are not particularly similar to any of the UShER barcode definitions, possibly indicating the presence of more lineages than what I had specified in ProVoC.

\subsection{Temporal Binomial NMF with Sum-to-Exactly-One (TBNMF=1)}

In this version of the TBNMF, there is still a latent spline representation of the abundance of each lineage, but now these splines act as prior distributions for a Dirichlet distribution.
The added Dirichlet allows for extra variation in the estimates of abundance.
In a Dirichlet distribution, the location parameter is a vector, often called $\underline\alpha$.
This parameter vector represents the abundance of each lineage at time $t$.
There is no restriction that $\sum_i\alpha_i=1$ as long as $\alpha_i > 0$ for all $\alpha_i$.
Higher values of $\alpha_j$ indicate lower variance: if $\alpha_j/\sum_i\alpha_i = 0.5$ and $\alpha_j = 1$, then the realization of the Dirichlet random variable will be close to 0.5, but with some variance; if $\alpha_j = 10$ instead, then the realization will be very close to 0.5.
Because of this, the latent splines do not need to be ``fluffmaxed'', and are allowed to take values larger than one.

\begin{figure}[ht!]
    \centering
\includegraphics[width=0.99\textwidth]{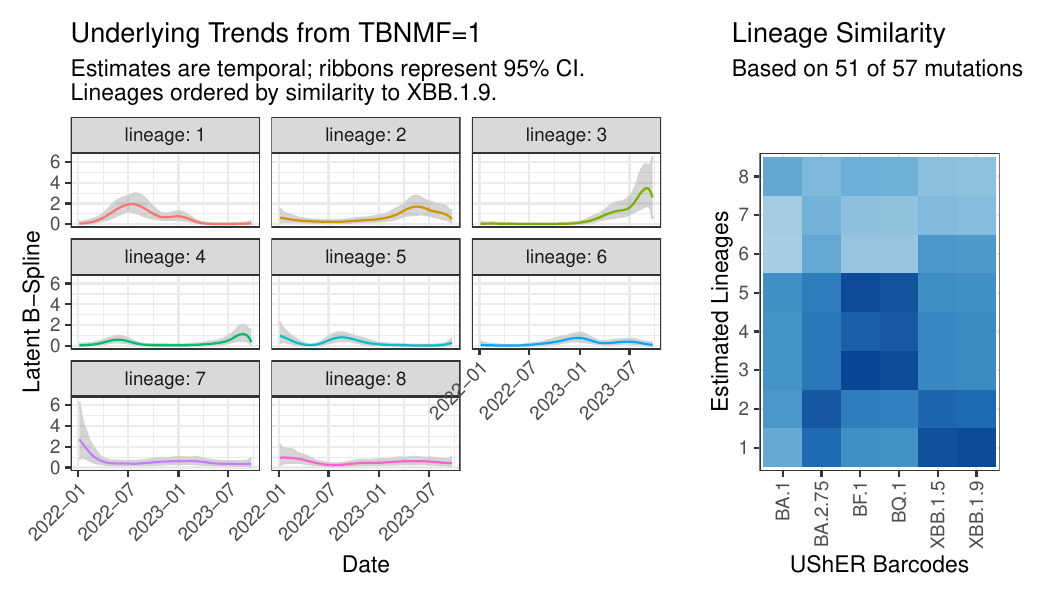}
\caption{\label{fig:unmuted_dp_latent}}
\end{figure}

\begin{figure}[ht!]
    \centering
\includegraphics[width=0.99\textwidth]{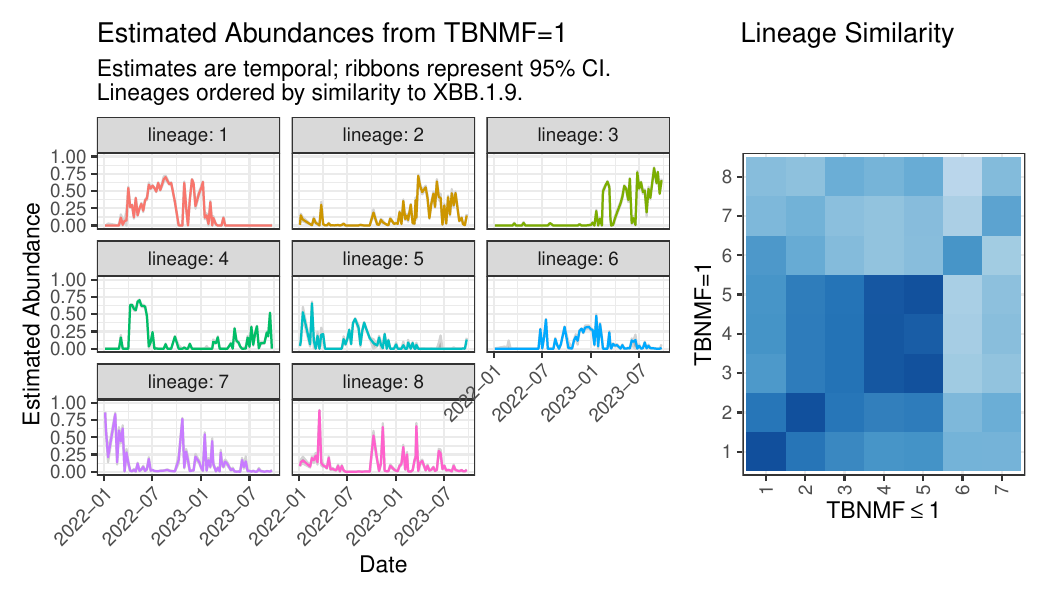}
\caption{\label{fig:unmuted_dp}}
\end{figure}

Figure \ref{fig:unmuted_dp_latent} shows the latent spline representation of the abundances along with a comparison of the lineages to the UShER barcodes, and Figure \ref{fig:unmuted_dp} shows the Dirichlet realizations along with a comparison of the lineages estimated by TBNMF$\le$1.
Many of the same patterns are evident as in TBNMF$\le$1, as is apparent in the lineage similarity plot in Figure \ref{fig:unmuted_dp}.
As before, the lineages are sorted by their similarity to XBB.1.9, which allows for direct comparison of lineages labelled ``1'', ``2'', etc.
The lineages labelled 1 from each method are quite similar between the two methods, as are the lineages labelled 2, 6, and 7.
There is high similarity between lineages 4 and 5 from TBNMF$\le$1 with lineages 3, 4, and 5 from TBNMF$=$1

However, this method allows for extra variation in the estimates of the abundance.
The spline term for lineage 3 has values near 4 towards the end of the sampling period, which indicates that the realizations will be less variable.
In contrast, the lineages labelled 4, 5, and 6 all have values less than 1, indicating that the realizations can be further from the abundances that would be implied by $\alpha_i/\sum_j\alpha_j$.

\subsection{Comparison of Estimated Lineages}

\subsubsection{Between Methods}

\begin{figure}
    \centering
    \includegraphics[width=\textwidth]{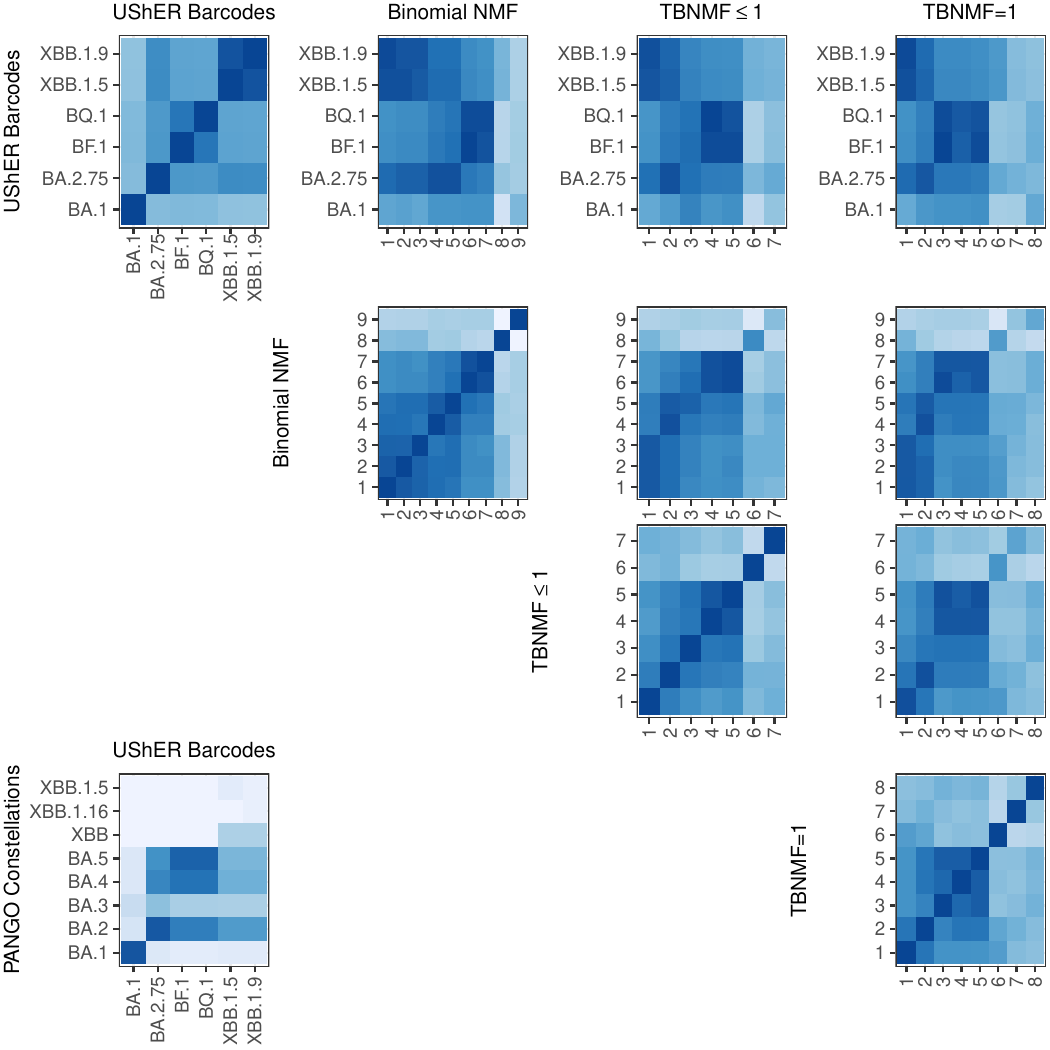}
    \caption{\label{fig:all-bases}A comparison between the estimated lineage definitions from each of the methods, as well as PANGO lineages and the ones used by Freyja (UShER barcodes). Darker colours indicate higher Jaccard similarity. The plot in the bottom left demonstrates the similarities and differences between PANGO lineages and UShER barcodes.}
\end{figure}

In this section I compare the estimated lineages from each method to each other as well as to those defined by PANGO and UShER.
Results are shown in Figure \ref{fig:all-bases}.
The main diagonal of the plot shows the similarities of lineages within a method, whereas the upper triangle shows the similarities of lineage definitions between methods.
The titles of the plots on the top row apply to all plots in that column, whereas the y axis labels apply to all plots in that row.
The notable exception to these rules is the plot at the very bottom left, which is the only plot that shows the lineage definitions used in the PANGO constellations.

There are several striking patterns in this plot.
Firstly, the UShER barcodes and PANGO constellations appear to have very little similarity. 
Note that the similarities are based on 55 mutations, even though the barcodes have 136 mutations relevant to this study and the constellations have 84 (only 55 mutations are shared between the definitions).
The PANGO constellations do not have definitions BA.2.75, BF.1, BQ.1, nor XBB.1.9, so I chose to display all BA.* lineages and also include XBB.1.16.
BA.1 appears to be consistent in both definitions, and BA.2.75 in the barcodes is similar to BA.2 in the constellations.
The constellations for BA.2, BA.4, and BA.5 have overlap with the barcodes for BA.2.75, BF.1, and BQ.1.
However, the recombinant definitions are completely different between the two (PANGO constellations stopped being updated around the time that the recombinants appeared).

\subsubsection{Between Locations}

The data in NCBI BioProject {PRJNA1088471} contains several locations and the methods in this paper were applied to each of those locations.
For Binomial NMF, the results are shown in Figure \ref{fig:loc_comparison}.
For this analysis, I chose to estimate 6 lineages in each location for consistency.

The main diagonal shows the similarity of mutations within a location, whereas the upper diagonal shows a pairwise comparison of estimated lineages between locations.
While the results aren't perfectly concordant (especially for Ashbridges Bay), each location resulted in lineages that were broadly similar.
For each plot in the upper triangle of plots, the diagonal from bottom left to top right shows the similarity between lineage 1 in both locations, where the lineages are a gain ordered by their similarity to XBB.1.9.
This diagonal is generally the highest similarity, with the occaisonal situation where two lineages from one location are similar to one lineage from another location (consistent with the similarity of BA.2.75, BF.1, and BQ.1).

\section{Conclusions and Future Work}

The results shown in this study clearly demonstrate that incorporating information across temporal samples allows for estimation of the definitions of lineages without requiring clinical samples.
The resultant lineage definitions are not perfectly aligned with known definitions but are sufficient to capture the overall dynamics (also, ``known definitions'' should not be treated as a ground truth).

The novel methods presented in this paper allow for mutations to be present in multiple estimated lineages, rather than using a strict clustering method which assigns each mutation to exactly one cluster.
This means that the methods can effectively estimate lineages and their derivative lineages, such as BA.2.75 and BQ.1, although these lineages are quite similar even in the known definitions and the exact mutations may difficult to differentiate.

In theory, the method should be able to set lineages to 0 to effectively remove them from the model, thus resulting in a similar set of lineages as long as the rank was set sufficiently high.
In practice, lineages were occaisonally set to 0, and this did not result in the same results.
The method for removing lineages was using an indicator variable for the temporal trends, which multiplies the whole trend by either 0 or 1.
An important consequence of this is that the model still tries to estimate the temporal trend, even though it will be multiplied by zero later in the estimation routine.
A more appropriate approach may be to use Reversible Jump Markov Chain Monte Carlo (RJMCMC), which allows for variables to be removed from the estimation routine entirely (or added).
However, software implementations of this act on a single variable and further research is needed to apply this to an entire time series.

There are several other potential improvements to this model.
For this particular application we have several locations, and a model that incorporates all locations (using the same lineage at all locations) would likely provide even better estimates. 
In general, there is no restriction for this model to use B-Splines; this simply provided a smooth temporal trend for the underlying abundances of lineages.
Penalized splines, Gaussian processes, and autoregressive models would all be appropriate for the temporal trend in abundances, although considerations will need to be made to satisfy the sum-to-one(-or-less) constraints.
The estimated lineage definitions allowed for a mutation's inclusion in a lineage to be completely independent from other mutations in that lineage, which in this case lead to lineage definitions with many mutations present and a lot of overlap.
It may be preferable to restrict a lineage to have fewer mutations, \emph{i.e.} fewer than half of the mutations are allowed to be present in a definition.
Finally, there were several aribtrary decisions made in the data processing (such as a choice of threshold for inclusion of mutations), and further investigation of these decisions would be required prior to using the estimated lineage definitions in practice.

Overall, the methods introduced in this paper are statistically sound unsupervised machine learning methods that are able to estimate the temporally smooth abundance of lineages while also estimating what those definitions are.
The estimated definitions provide an alternative to the use of clinical data, which is becoming rarer as the pandemic wanes.
The methods in this paper are not specific to SARS-CoV-2, and can easily be applied mpox, influenza, etc. with only a modification to the data pre-processing.

\section*{Acknowledgements}

This work was generously funded by NSERC RGPIN-2023-05534. I gratefully acknowledge the discussions and comments on this work by Jennifer Knapp, Alyssa Overton, and Trevor Charles.

\bibliographystyle{apalike}
\bibliography{ms-unmuted.bbl}

\clearpage

\section*{Appendix}

\appendix

\renewcommand{\thefigure}{S\arabic{figure}}
\setcounter{figure}{0}

\begin{figure}[ht!]
    \centering
    \includegraphics[width=0.95\textwidth]{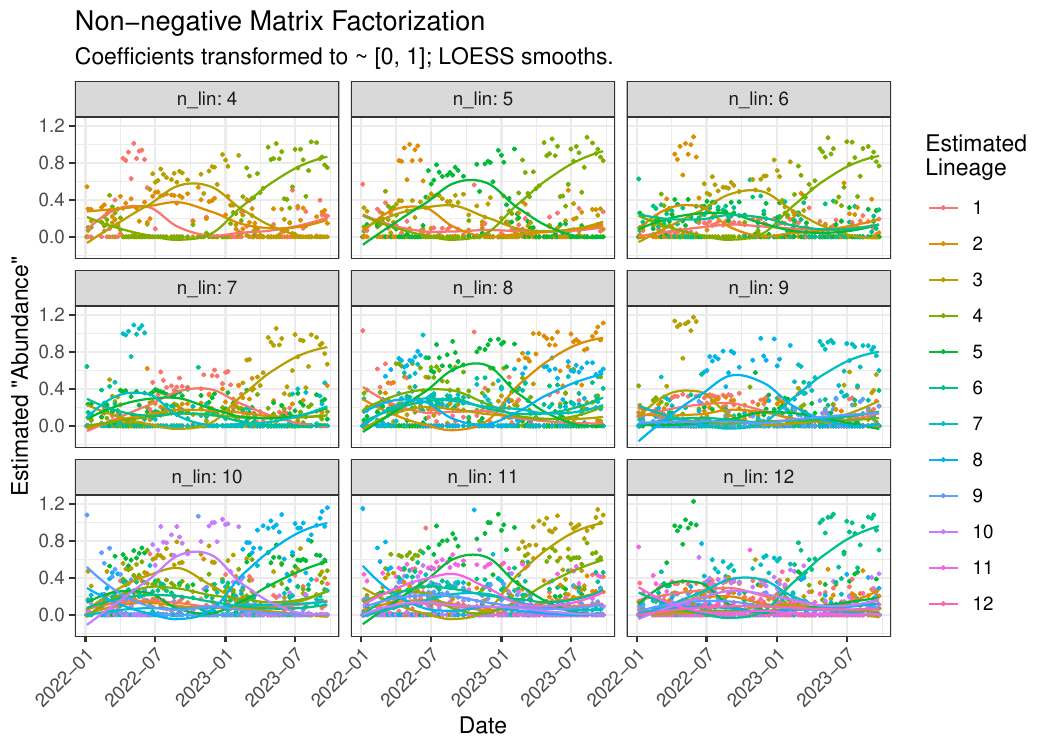}
    \caption{\label{fig:nmf_all}Coefficients representing the prevalence of each estimated lineage from NMF, fit separately to each location. The lineage names are arbitrary; they are not based on placement in a phylogenetic tree as defined by clinical cases. In all cases, the results approximately match the results from when the lineages were known from clinical cases.}
\end{figure}

\begin{figure}[ht!]
    \centering
    \includegraphics[width=0.95\textwidth]{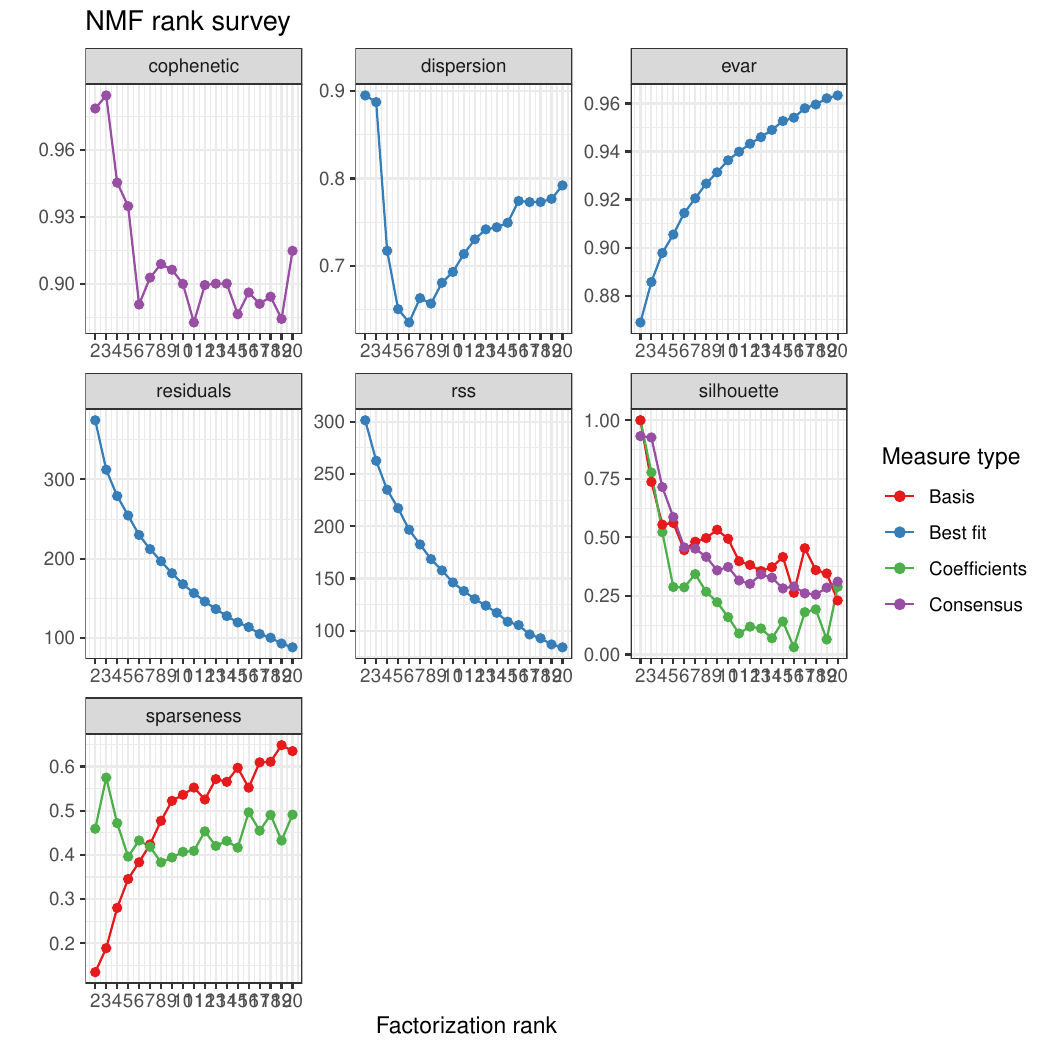}
    \caption{\label{fig:nmf_rank}Results of fitting NMF for ranks of 2 to 20.}
\end{figure}

\begin{figure}[ht!]
    \centering
    \includegraphics[width=0.95\textwidth]{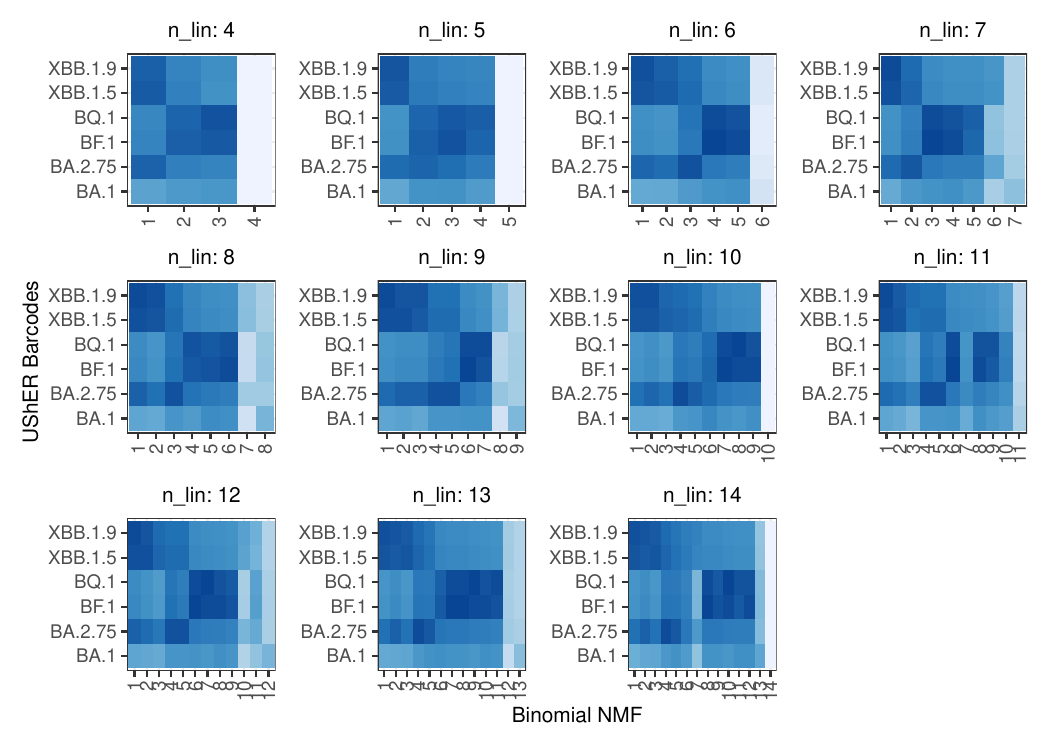}
    \caption{\label{fig:bbnmf_rank}An investigation of the results for various ranks in the Binomial NMF version of the model. By estimating more and more lineages, the algorithm appears to split a combination BA.2.75, XBB.1.9, and XBB.1.5 into more and more different lineages, and similar for BQ.1 and BF.1.}
\end{figure}

\begin{figure}[ht!]
    \centering
    \includegraphics[width=0.95\textwidth]{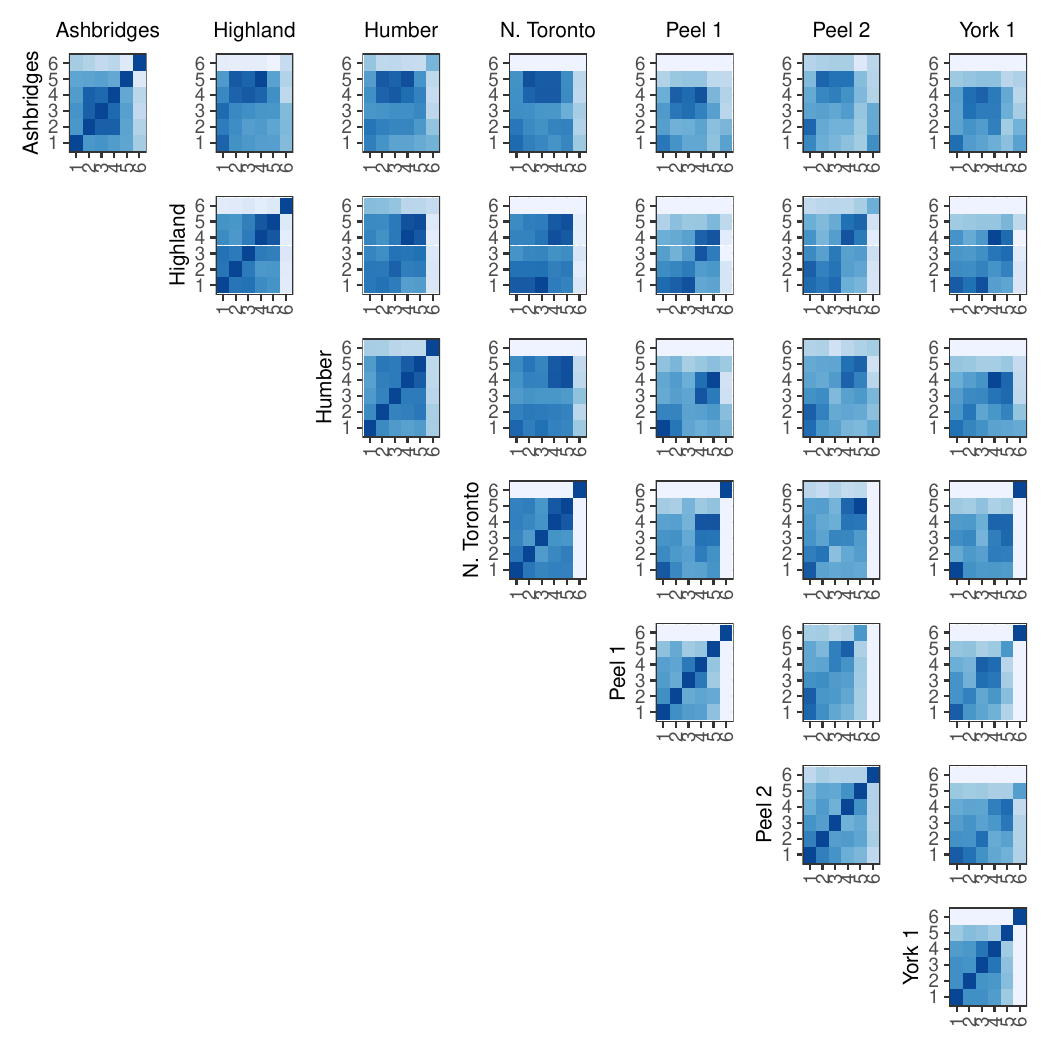}
    \caption{\label{fig:loc_comparison}Pairwise comparison of the lineage estimated from each location in the Binomial NMF model. For consistency, six lineages were estimated at each location. The main diagonal of plots shows the Jaccard similarities of lineages within a location, providing context for how lineages were defined by the algorithm (\emph{e.g.}, lineages 4 and 5 in Highland Creek were similar to each other, so it's not surprising that both are similar to multiple lineages in other locations). The upper triangle of plots shows the pairwise comparison of lineages between locations. In most situations, each plot has the highest similarity between lineages with the same number (\emph{e.g.}, lineage 1 is similar to lineage 1, lineage 2 is similar to lineage 2, etc.), noting that lineage numbers are ordered according to similarity to XBB.1.9.}
\end{figure}

\begin{figure}[ht!]
    \centering
    \includegraphics[width=0.95\textwidth]{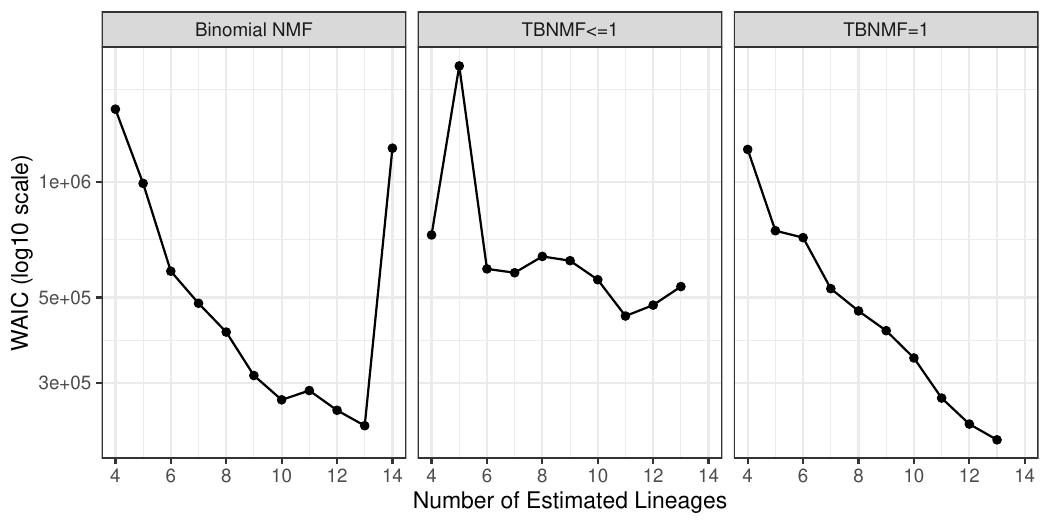}
    \caption{\label{fig:waic}WAIC results for each model. The WAIC for Binomial NMF has local minima at 10 and 13, TBNMF$\le$1 has local minima at 6, 7, and 11, and TBNMF=1 appears to continue decreasing, but with a large drop between 6 and 7 lineages. Based on the plots of the results and from domain knowledge, it is unreasonable to have more than 13 lineages. }
\end{figure}

\clearpage

\subsection*{Estimated Lineage Definitions}

\begin{table}[ht]
\centering
\small
\begin{tabular}{rrrrrrrrrr}
  \hline
 & 1 & 2 & 3 & 4 & 5 & 6 & 7 & 8 & 9 \\ 
  \hline
A27259C & 1 & 1 & 1 & 1 & 1 & 0 & 0 & 0 & 1 \\ 
  C10198T & 1 & 1 & 1 & 1 & 1 & 1 & 1 & 1 & 0 \\ 
  C12880T & 1 & 1 & 1 & 1 & 1 & 1 & 1 & 1 & 0 \\ 
  C15738T & 1 & 0 & 0 & 0 & 0 & 0 & 0 & 1 & 0 \\ 
  C25000T & 1 & 1 & 1 & 1 & 1 & 1 & 1 & 0 & 1 \\ 
  C25416T & 1 & 1 & 1 & 1 & 0 & 0 & 0 & 1 & 0 \\ 
  C25584T & 1 & 1 & 1 & 1 & 1 & 1 & 1 & 0 & 1 \\ 
  C26858T & 1 & 1 & 1 & 1 & 1 & 0 & 0 & 1 & 0 \\ 
  C27807T & 1 & 1 & 1 & 1 & 1 & 1 & 1 & 1 & 0 \\ 
  C27889T & 0 & 0 & 0 & 0 & 0 & 1 & 1 & 0 & 0 \\ 
  C3037T & 1 & 1 & 1 & 1 & 1 & 1 & 1 & 0 & 0 \\ 
  G10447A & 1 & 1 & 1 & 1 & 1 & 1 & 1 & 1 & 0 \\ 
  G12160A & 0 & 0 & 0 & 0 & 0 & 1 & 1 & 0 & 0 \\ 
  G28882A & 1 & 1 & 1 & 1 & 1 & 1 & 1 & 1 & 0 \\ 
  T15939C & 1 & 1 & 1 & 0 & 0 & 0 & 0 & 1 & 0 \\ 
  T17859C & 1 & 1 & 1 & 0 & 0 & 0 & 0 & 1 & 0 \\ 
  T27384C & 1 & 1 & 1 & 1 & 1 & 0 & 0 & 1 & 0 \\ 
  T28297C & 1 & 0 & 1 & 0 & 0 & 0 & 0 & 0 & 0 \\ 
  aa:E:T11A & 1 & 1 & 1 & 0 & 0 & 0 & 0 & 1 & 0 \\ 
  aa:E:T9I & 1 & 1 & 1 & 1 & 1 & 1 & 1 & 1 & 0 \\ 
  aa:M:A63T & 1 & 1 & 1 & 1 & 1 & 1 & 1 & 1 & 0 \\ 
  aa:N:E136D & 0 & 0 & 0 & 0 & 0 & 0 & 1 & 0 & 0 \\ 
  aa:N:G204R & 1 & 1 & 1 & 1 & 1 & 1 & 1 & 0 & 1 \\ 
  aa:N:R203K & 1 & 1 & 1 & 1 & 1 & 1 & 1 & 0 & 0 \\ 
  aa:N:S413R & 1 & 1 & 1 & 1 & 1 & 1 & 1 & 0 & 0 \\ 
  aa:S:D614G & 1 & 1 & 1 & 1 & 1 & 1 & 1 & 0 & 1 \\ 
  aa:S:D796Y & 1 & 1 & 1 & 1 & 1 & 1 & 1 & 0 & 0 \\ 
  aa:S:E484A & 1 & 1 & 1 & 1 & 1 & 1 & 1 & 0 & 0 \\ 
  aa:S:G142D & 1 & 1 & 0 & 1 & 1 & 1 & 1 & 0 & 0 \\ 
  aa:S:H655Y & 1 & 1 & 1 & 1 & 1 & 1 & 1 & 0 & 1 \\ 
  aa:S:N501Y & 1 & 1 & 1 & 1 & 1 & 1 & 1 & 0 & 1 \\ 
  aa:S:N679K & 1 & 1 & 1 & 1 & 1 & 1 & 1 & 0 & 1 \\ 
  aa:S:N969K & 1 & 1 & 1 & 1 & 1 & 1 & 1 & 1 & 0 \\ 
  aa:S:P681H & 1 & 1 & 1 & 1 & 1 & 1 & 1 & 0 & 1 \\ 
  aa:S:Q498R & 1 & 1 & 1 & 1 & 1 & 1 & 1 & 0 & 0 \\ 
  aa:S:Q954H & 1 & 1 & 1 & 1 & 1 & 1 & 1 & 0 & 1 \\ 
  aa:S:S477N & 1 & 1 & 1 & 1 & 1 & 1 & 1 & 0 & 0 \\ 
  aa:S:T19I & 1 & 1 & 1 & 1 & 1 & 1 & 1 & 0 & 1 \\ 
  aa:S:T478K & 1 & 1 & 0 & 1 & 1 & 1 & 1 & 0 & 0 \\ 
  aa:S:Y505H & 1 & 1 & 1 & 1 & 1 & 1 & 1 & 0 & 0 \\ 
  aa:orf1a:G1307S & 1 & 1 & 1 & 1 & 1 & 1 & 1 & 0 & 0 \\ 
  aa:orf1a:K47R & 1 & 1 & 1 & 0 & 0 & 0 & 0 & 1 & 0 \\ 
  aa:orf1a:P3395H & 1 & 1 & 1 & 1 & 1 & 1 & 1 & 0 & 1 \\ 
  aa:orf1a:S135R & 0 & 0 & 1 & 0 & 1 & 1 & 0 & 1 & 0 \\ 
  aa:orf1a:T3090I & 1 & 1 & 1 & 1 & 1 & 1 & 1 & 1 & 0 \\ 
  aa:orf1a:T3255I & 1 & 1 & 1 & 1 & 1 & 1 & 1 & 0 & 0 \\ 
  aa:orf1a:T842I & 1 & 1 & 1 & 1 & 1 & 1 & 1 & 0 & 0 \\ 
  aa:orf1b:P314L & 1 & 1 & 1 & 1 & 1 & 1 & 1 & 0 & 1 \\ 
  aa:orf1b:R1315C & 1 & 1 & 1 & 1 & 1 & 1 & 1 & 1 & 0 \\ 
  aa:orf6:D61H & 1 & 1 & 1 & 1 & 1 & 0 & 0 & 1 & 0 \\ 
  aa:orf6:D61V & 1 & 1 & 1 & 1 & 1 & 0 & 0 & 1 & 0 \\ 
  aa:orf8:G8* & 1 & 1 & 0 & 0 & 0 & 0 & 0 & 1 & 0 \\ 
  del:11288:9 & 1 & 1 & 1 & 1 & 1 & 1 & 1 & 1 & 0 \\ 
  del:21633:9 & 1 & 0 & 1 & 1 & 1 & 1 & 1 & 0 & 0 \\ 
  del:21765:6 & 0 & 0 & 0 & 0 & 0 & 1 & 1 & 0 & 1 \\ 
  del:28362:9 & 0 & 1 & 1 & 0 & 1 & 1 & 1 & 1 & 0 \\ 
  del:29734:26 & 1 & 1 & 1 & 0 & 1 & 1 & 1 & 0 & 0 \\ 
   \hline
\end{tabular}
\caption{\label{tab:bbnmf_w}Estimated lineage definitions from BNMF}
\end{table}

\begin{table}[ht]
\centering
\small
\begin{tabular}{rrrrrrrr}
  \hline
 & 1 & 2 & 3 & 4 & 5 & 6 & 7 \\ 
  \hline
A27259C & 1 & 1 & 1 & 0 & 0 & 1 & 1 \\ 
  C10198T & 1 & 1 & 1 & 1 & 1 & 1 & 0 \\ 
  C12880T & 1 & 1 & 1 & 1 & 1 & 1 & 0 \\ 
  C15738T & 1 & 0 & 0 & 0 & 0 & 0 & 0 \\ 
  C25000T & 1 & 1 & 1 & 1 & 1 & 0 & 1 \\ 
  C25416T & 1 & 1 & 0 & 0 & 0 & 1 & 0 \\ 
  C25584T & 1 & 1 & 1 & 1 & 1 & 0 & 1 \\ 
  C26858T & 1 & 1 & 1 & 0 & 0 & 1 & 1 \\ 
  C27807T & 1 & 1 & 1 & 1 & 1 & 1 & 0 \\ 
  C27889T & 0 & 0 & 0 & 1 & 1 & 0 & 0 \\ 
  C3037T & 1 & 1 & 1 & 1 & 1 & 0 & 0 \\ 
  G10447A & 1 & 1 & 0 & 1 & 1 & 1 & 1 \\ 
  G12160A & 0 & 0 & 0 & 1 & 1 & 0 & 0 \\ 
  G28882A & 1 & 1 & 1 & 1 & 1 & 1 & 0 \\ 
  T15939C & 1 & 0 & 0 & 0 & 0 & 1 & 0 \\ 
  T17859C & 1 & 0 & 0 & 0 & 0 & 1 & 0 \\ 
  T27384C & 1 & 1 & 0 & 0 & 0 & 1 & 1 \\ 
  T28297C & 1 & 0 & 0 & 0 & 0 & 0 & 0 \\ 
  aa:E:T11A & 1 & 0 & 0 & 0 & 0 & 1 & 0 \\ 
  aa:E:T9I & 1 & 1 & 1 & 1 & 1 & 1 & 0 \\ 
  aa:M:A63T & 1 & 1 & 1 & 1 & 1 & 1 & 0 \\ 
  aa:N:E136D & 0 & 0 & 0 & 1 & 0 & 0 & 0 \\ 
  aa:N:G204R & 1 & 1 & 1 & 1 & 1 & 1 & 0 \\ 
  aa:N:R203K & 1 & 1 & 1 & 1 & 1 & 0 & 0 \\ 
  aa:N:S413R & 1 & 1 & 1 & 1 & 1 & 0 & 1 \\ 
  aa:S:D614G & 1 & 1 & 1 & 1 & 1 & 0 & 1 \\ 
  aa:S:D796Y & 1 & 1 & 1 & 1 & 1 & 0 & 0 \\ 
  aa:S:E484A & 1 & 1 & 1 & 1 & 1 & 0 & 0 \\ 
  aa:S:G142D & 0 & 1 & 1 & 1 & 1 & 1 & 0 \\ 
  aa:S:H655Y & 1 & 1 & 1 & 1 & 1 & 0 & 1 \\ 
  aa:S:N501Y & 1 & 1 & 1 & 1 & 1 & 0 & 1 \\ 
  aa:S:N679K & 1 & 1 & 1 & 1 & 1 & 0 & 1 \\ 
  aa:S:N969K & 1 & 1 & 1 & 1 & 1 & 0 & 1 \\ 
  aa:S:P681H & 1 & 1 & 1 & 1 & 1 & 1 & 0 \\ 
  aa:S:Q498R & 1 & 1 & 1 & 1 & 1 & 0 & 1 \\ 
  aa:S:Q954H & 1 & 1 & 1 & 1 & 1 & 0 & 1 \\ 
  aa:S:S477N & 1 & 1 & 1 & 1 & 1 & 0 & 0 \\ 
  aa:S:T19I & 1 & 1 & 1 & 1 & 1 & 0 & 1 \\ 
  aa:S:T478K & 0 & 1 & 1 & 1 & 1 & 1 & 0 \\ 
  aa:S:Y505H & 1 & 1 & 1 & 1 & 1 & 0 & 1 \\ 
  aa:orf1a:G1307S & 1 & 1 & 1 & 1 & 1 & 0 & 0 \\ 
  aa:orf1a:K47R & 1 & 1 & 0 & 0 & 0 & 1 & 0 \\ 
  aa:orf1a:P3395H & 1 & 1 & 1 & 1 & 1 & 0 & 1 \\ 
  aa:orf1a:S135R & 0 & 1 & 1 & 1 & 0 & 1 & 0 \\ 
  aa:orf1a:T3090I & 1 & 1 & 0 & 1 & 1 & 1 & 1 \\ 
  aa:orf1a:T3255I & 1 & 1 & 1 & 1 & 1 & 0 & 0 \\ 
  aa:orf1a:T842I & 1 & 1 & 1 & 1 & 1 & 0 & 1 \\ 
  aa:orf1b:P314L & 1 & 1 & 1 & 1 & 1 & 0 & 1 \\ 
  aa:orf1b:R1315C & 1 & 1 & 1 & 1 & 1 & 1 & 0 \\ 
  aa:orf6:D61H & 1 & 1 & 0 & 0 & 0 & 1 & 1 \\ 
  aa:orf6:D61V & 1 & 1 & 0 & 0 & 0 & 1 & 1 \\ 
  aa:orf8:G8* & 1 & 0 & 0 & 0 & 0 & 0 & 0 \\ 
  del:11288:9 & 1 & 1 & 1 & 1 & 1 & 1 & 0 \\ 
  del:21633:9 & 0 & 1 & 1 & 1 & 1 & 1 & 0 \\ 
  del:21765:6 & 0 & 0 & 0 & 1 & 1 & 0 & 0 \\ 
  del:28362:9 & 1 & 0 & 1 & 0 & 1 & 1 & 1 \\ 
  del:29734:26 & 1 & 0 & 1 & 1 & 1 & 0 & 1 \\ 
   \hline
\end{tabular}
\caption{\label{tab:unmuted_w}Estimated lineage definitions from TBNMF$\le$1}
\end{table}

\begin{table}[ht]
\centering
\small
\begin{tabular}{rrrrrrrrr}
  \hline
 & 1 & 2 & 3 & 4 & 5 & 6 & 7 & 8 \\ 
  \hline
A27259C & 1 & 1 & 0 & 0 & 0 & 1 & 1 & 1 \\ 
  C10198T & 1 & 1 & 1 & 1 & 1 & 1 & 1 & 0 \\ 
  C12880T & 1 & 1 & 1 & 1 & 1 & 1 & 0 & 0 \\ 
  C15738T & 1 & 0 & 0 & 0 & 0 & 1 & 0 & 0 \\ 
  C25000T & 1 & 1 & 1 & 1 & 1 & 0 & 1 & 1 \\ 
  C25416T & 1 & 1 & 0 & 0 & 0 & 1 & 0 & 0 \\ 
  C25584T & 1 & 1 & 1 & 1 & 1 & 1 & 0 & 1 \\ 
  C26858T & 1 & 1 & 0 & 1 & 0 & 1 & 0 & 0 \\ 
  C27807T & 1 & 1 & 1 & 1 & 1 & 1 & 0 & 1 \\ 
  C27889T & 0 & 0 & 1 & 0 & 1 & 0 & 0 & 1 \\ 
  C3037T & 1 & 1 & 1 & 1 & 1 & 0 & 1 & 0 \\ 
  G10447A & 1 & 1 & 1 & 1 & 1 & 1 & 1 & 0 \\ 
  G12160A & 0 & 0 & 1 & 1 & 1 & 0 & 0 & 0 \\ 
  G28882A & 1 & 1 & 1 & 1 & 1 & 1 & 0 & 1 \\ 
  T15939C & 1 & 0 & 0 & 0 & 0 & 1 & 0 & 0 \\ 
  T17859C & 1 & 0 & 0 & 0 & 0 & 1 & 0 & 0 \\ 
  T27384C & 1 & 1 & 0 & 0 & 0 & 1 & 1 & 0 \\ 
  T28297C & 1 & 0 & 0 & 0 & 0 & 0 & 0 & 0 \\ 
  aa:E:T11A & 1 & 0 & 0 & 0 & 0 & 1 & 0 & 0 \\ 
  aa:E:T9I & 1 & 1 & 1 & 1 & 1 & 1 & 0 & 1 \\ 
  aa:M:A63T & 1 & 1 & 1 & 1 & 1 & 1 & 0 & 1 \\ 
  aa:N:E136D & 0 & 0 & 0 & 1 & 0 & 0 & 0 & 0 \\ 
  aa:N:G204R & 1 & 1 & 1 & 1 & 1 & 1 & 0 & 1 \\ 
  aa:N:R203K & 1 & 1 & 1 & 1 & 1 & 1 & 0 & 0 \\ 
  aa:N:S413R & 1 & 1 & 1 & 1 & 1 & 1 & 0 & 1 \\ 
  aa:S:D614G & 1 & 1 & 1 & 1 & 1 & 0 & 1 & 1 \\ 
  aa:S:D796Y & 1 & 1 & 1 & 1 & 1 & 0 & 0 & 0 \\ 
  aa:S:E484A & 1 & 1 & 1 & 1 & 1 & 0 & 0 & 0 \\ 
  aa:S:G142D & 1 & 1 & 1 & 1 & 1 & 0 & 0 & 0 \\ 
  aa:S:H655Y & 1 & 1 & 1 & 1 & 1 & 0 & 1 & 1 \\ 
  aa:S:N501Y & 1 & 1 & 1 & 1 & 1 & 1 & 0 & 0 \\ 
  aa:S:N679K & 1 & 1 & 1 & 1 & 1 & 0 & 1 & 1 \\ 
  aa:S:N969K & 1 & 1 & 1 & 1 & 1 & 0 & 1 & 1 \\ 
  aa:S:P681H & 1 & 1 & 1 & 1 & 1 & 0 & 1 & 1 \\ 
  aa:S:Q498R & 1 & 1 & 1 & 1 & 1 & 1 & 0 & 0 \\ 
  aa:S:Q954H & 1 & 1 & 1 & 1 & 1 & 0 & 1 & 1 \\ 
  aa:S:S477N & 1 & 1 & 1 & 1 & 1 & 0 & 0 & 0 \\ 
  aa:S:T19I & 1 & 1 & 1 & 1 & 1 & 0 & 1 & 1 \\ 
  aa:S:T478K & 0 & 1 & 1 & 1 & 1 & 1 & 0 & 0 \\ 
  aa:S:Y505H & 1 & 1 & 1 & 1 & 1 & 1 & 0 & 0 \\ 
  aa:orf1a:G1307S & 1 & 1 & 1 & 1 & 1 & 0 & 1 & 0 \\ 
  aa:orf1a:K47R & 1 & 1 & 0 & 0 & 0 & 1 & 0 & 0 \\ 
  aa:orf1a:P3395H & 1 & 1 & 1 & 1 & 1 & 0 & 1 & 1 \\ 
  aa:orf1a:S135R & 1 & 0 & 1 & 0 & 0 & 0 & 0 & 0 \\ 
  aa:orf1a:T3090I & 1 & 1 & 1 & 1 & 1 & 1 & 1 & 0 \\ 
  aa:orf1a:T3255I & 1 & 1 & 1 & 1 & 1 & 1 & 0 & 0 \\ 
  aa:orf1a:T842I & 1 & 1 & 1 & 1 & 1 & 0 & 1 & 0 \\ 
  aa:orf1b:P314L & 1 & 1 & 1 & 1 & 1 & 0 & 1 & 1 \\ 
  aa:orf1b:R1315C & 1 & 1 & 1 & 1 & 1 & 1 & 0 & 1 \\ 
  aa:orf6:D61H & 1 & 1 & 0 & 0 & 0 & 1 & 1 & 0 \\ 
  aa:orf6:D61V & 1 & 1 & 0 & 0 & 0 & 1 & 1 & 0 \\ 
  aa:orf8:G8* & 1 & 0 & 0 & 0 & 0 & 1 & 0 & 0 \\ 
  del:11288:9 & 1 & 1 & 1 & 1 & 1 & 1 & 1 & 0 \\ 
  del:21633:9 & 0 & 1 & 1 & 1 & 1 & 1 & 1 & 0 \\ 
  del:21765:6 & 0 & 0 & 1 & 1 & 0 & 0 & 0 & 1 \\ 
  del:28362:9 & 1 & 0 & 1 & 0 & 0 & 1 & 1 & 1 \\ 
  del:29734:26 & 1 & 1 & 0 & 1 & 1 & 1 & 0 & 1 \\ 
   \hline
\end{tabular}
\caption{\label{tab:unmuted_dp_w}Estimated lineage definitions from TBNMF=1}
\end{table}

\end{document}